\documentclass[pra,reprint,showpacs,amsmath,amssymb,superscriptaddress]{revtex4-2}
\usepackage{graphicx}
\usepackage{amsmath}
\usepackage{bm}
\usepackage{amssymb}
\usepackage{soul}
 \usepackage[version=3]{mhchem}
\usepackage[colorlinks,bookmarks=false,citecolor=blue,linkcolor=red,urlcolor=blue]{hyperref}
\usepackage{esint}
\usepackage{dsfont}
\usepackage[bottom]{footmisc}
\usepackage{physics}
\usepackage[normalem]{ulem} % to support \sout{text goes here} command
\usepackage{mathtools}
\usepackage{footnote}

\newcommand{\be}{\begin{equation}}
\newcommand{\ee}{\end{equation}}
\newcommand{\bea}{\begin{eqnarray}}
\newcommand{\eea}{\end{eqnarray}}

\begin{document}
\title{Field-induced Peierls phase in $S=1$ Heisenberg spins coupled to quantum phonons}
\author{Shifeng Cui}
\affiliation{Shandong Engineering Research Center of Aeronautical Materials and Devices, College of Aeronautical Engineering, Shandong University of Aeronautics, Binzhou, 256603, China}
\affiliation{Beijing Computational Science Research Center, Beijing 100193, China}
\author{Wenan Guo}
\email{waguo@bnu.edu.cn}
\affiliation{Department of Physics, Beijing Normal University, Beijing 100875, China}
\affiliation{Key Laboratory of Multiscale Spin Physics (Ministry of Education), Beijing Normal University, Beijing 100875, China}
\author{G. G. Batrouni}
\email{george.batrouni@univ-cotedazur.fr}
\affiliation{Universit\'e C\^ote d'Azur, CNRS, INPHYNI, France}
\author{Pinaki Sengupta}
\email{psengupta@ntu.edu.sg}
\affiliation{Nanyang Technological University, Singapore, Singapore}

\begin{abstract}
Spin-Peierls transition occurs in a one-dimensional $S=1$ Heisenberg antiferromagnetic model 
with single-ion anisotropy, coupled to finite frequency bond phonons, in a magnetic field. Our 
results indicate that for the pure Heisenberg model, any Peierls transition is suppressed by quantum
fluctuations of the phonon field. However, a novel magnetic field-induced Spin-Peierls phase is realized in 
the presence of strong single-ion anisotropy. Contrary to the standard Peierls state,  the periodicity 
of bond strength modulation in this field-induced Spin-Peierls state is variable and depends 
on the strength of the applied field. The nature of the ground state in this new phase and the 
associated field-driven transitions to and out of this phase are explored using extensive 
numerical simulations. In particular, we explore the spin and bond correlations and the evolution of bond order modulation with varying magnetic field.

\end{abstract}


\maketitle

\section{Introduction}

Interaction with lattice degrees of freedom results in novel electronic 
states in matter. In the 1950s, Frohlich\cite{frohlich54} and 
Peierls\cite{peierls55} showed that non-interacting electrons in one 
dimension (1D), coupled to an elastic lattice, will experience an 
instability towards lattice distortion. For a half-filled band 
(one electron per site), this leads to a spontaneous dimerization 
of the 1D chain (Peierls state). Later studies established that 
the Peierls transition persists even in the presence of 
electron-electron interactions. Experimentally, 
the Peierls instability is manifested in several quasi-1D materials, 
e.g., conjugated polymers, organic charge transfer salts, MX salts, 
etc. In quantum spin systems (Mott insulators with frozen charge 
degrees of freedom), an analogous spin-Peierls (SP) transition 
is observed. As a paradigmatic example, the spin-${1\over 2}$ 
Heisenberg antiferromagnetic chain exhibits instability to lattice 
dimerization. The elastic energy cost ($\sim\delta^2$) due to 
lattice distortion $\delta$  is smaller than the associated gain 
in exchange energy ($\sim\delta^{4/3}$). This stabilizes a ground 
state with nonzero dimerization and a spin gap. In the adiabatic 
limit (zero frequency, classical phonons), any arbitrarily small 
spin-phonon coupling leads to a dimerized ground state in a spin 
chain. On the other hand, for finite frequency phonons, quantum 
lattice fluctuations suppress bond distortion at small spin-phonon 
couplings and/or large phonon frequencies. Transition to a 
dimerized spin-Peierls state occurs for finite, nonzero critical 
spin-phonon coupling (that depends on the phonon frequency).
Experimentally, the spin-Peierls instability can be widely observed 
in quasi$-$1D materials, for example, organic polymer material 
TTF\cite{jacobs74}, metal oxide material CuGeO3\cite{hase93}, 
nitrate Tetrathiafulvalene salt\cite{soriano23}, and organic 
spin chain (o-DMTTF) 2X (X=Cl, Br, I)\cite{su79}, etc.

Compared to $S={1\over 2}$ spins{\color{red}{\cite{bursill99, aits03, weber20a}}}, the spin-Peierls transition has 
remained relatively less explored for $S>{1\over 2}$ spin systems{\color{red}\cite{guo90}}. 
Higher spin systems offer the potential to realize novel phases not realized in their 
$S={1\over 2}$ counterparts, due to the larger Hilbert space and 
additional interactions. Moreover, the ground state of $S=1$ 
antiferromagnetic (AFM) Heisenberg chain is gapped and non-magnetic, in contrast to that of the  $S={1\over 2}$ 
chain which is gapless and critical with quasi-long-range N{\' e}el 
AFM order{\color{red}{\cite{haldane83}}}, and any Spin-Peierls transition
will be distinct from that for $S={1\over 2}$ spins.

The spin-Peierls transition for $S=1$ Heisenberg AFM chain coupled to classical (zero frequency) 
phonons has been studied in detail using analytical and numerical methods
{\color{red}{\cite{pati96, onishi01, guang14}}}.
The transition to a dimerized spin-Peierls state is found to occur at a finite, nonzero
critical spin-phonon coupling strength, in contrast to that for 
$S={1\over 2}$ spins. However, the role of quantum (finite frequency) phonons has remained 
unexplored. Do quantum fluctuations (of phonons) 
move the spin-Peierls transition to higher values of the coupling strength
(as for $S={1\over 2}$ spins), or suppress it completely? What new
phases arise in the presence of additional interactions, such as
single-ion anisotropy that is ubiquitous in many $S=1$ quantum 
magnets? Incidentally, there have been far fewer experimental realizations 
of $S=1$ spin-Peierls transitions. It was initially believed that the ground 
state of \ce{LiVGe_2O_6} is a spin-1 Peierls state~\cite{millet99,gavilano00}, 
but subsequent studies showed that it lies in the AFM phase\cite{lumsden00}.
Recently, it has been proposed that there may be a Spin-1 Spin-Peierls 
phase transition in verdazyl-based salts\cite{yamaguchi23}.

Here, we present the results of our investigations of the ground state phases of a Spin-1 AFM Heisenberg chain with additional single-ion anisotropy, coupled to finite-frequency phonons in a
longitudinal magnetic field. We have used Density Matrix Renormalization Group (DMRG){\color{red}{\cite{white92}}}-based simulation of the model on finite-sized chains{\color{red}\cite{schollwock11}} to study the ground state phases in the different parameter 
regimes, searching for evidence of Peierls instability. Our results can be summarized as: 
(i) for the pure Heisenberg model (no single-ion anisotropy) in zero field, quantum phonons completely suppress the transition to a Peierls state for any physically relevant strength of spin-phonon coupling, 
(ii) for easy-axis single-ion anisotropy, the ground state exhibits lattice distortion induced by Peierls instability over a finite range of applied field, 
and (iii) the lattice modulation in the Peierls state varies with the applied field strength. Our results will be useful to experimentalists in their search for Peierls distortion in Spin-1 quantum magnets in low dimensions.

\section{Model and method} 
We study the $S=1$ AFM SP model with single-ion anisotropy 
coupled to bond phonons in a longitudinal magnetic field. The system is governed by the Hamiltonian, 
\begin{eqnarray}
\nonumber
H_{SP}&=&\sum_{i=1}^L \left [ J + \alpha \left ( {\hat a}^{\dagger}_i + {\hat a}^{\phantom\dagger}_i \right ) \right ]  {\hat S}_i \cdot {\hat S}_{i+1}  + \sum_{i=1}^L \omega{\hat a}^\dagger_i {\hat a}^{\phantom\dagger}_i  \\ 
&&+ \sum_{i=1}^L D (S_i^z)^2  + h \sum_{i=1}^L S_i^z,
\label{eq:sp-ham}
\end{eqnarray}
where $L$ is the number of sites, $J$ is the exchange interaction, $D$ is the single-ion anisotropy, and ${\hat a}^{\phantom\dagger}_i$ (${\hat a}^\dagger_i$) is the phonon destruction (creation) operator on the $i$th bond. The spin-phonon coupling strength is denoted by the parameter $\alpha$. The phonon frequency is $\omega$, and the strength of the longitudinal magnetic field is $h$. We set $J=1$ to fix the energy scale.

To characterize possible spin ordering, we calculate Green's function (transverse spin correlation function),
\begin{equation}
 C_{X,Y}(r) \equiv \frac{1}{L} \sum_i \langle S^+_i S_{i+r}^- + 
{\rm h.c.} \rangle.
\label{cxy}
\end{equation}
Power law decay of Green's function would indicate a quasi-long-range order for the 
transverse spin correlation. If Green's function decays to a finite constant, it signals 
long-range order and
the spontaneous breaking of the corresponding symmetry. The Fourier
transform of the transverse spin correlation function,
Eq.(\ref{cxy}), identifies the ordering wave vector. 

To identify any Peierls state, we calculate the bond-bond correlation,
\begin{eqnarray}
 D(r)\equiv \frac{1}{L} \sum_i \langle D(i)D(i+r) \rangle,
 \label{eq:dij}
\end{eqnarray}
where $D(i)={\hat S}_i \cdot {\hat S}_{i+1}$.
If the bond-bond correlation oscillates periodically, it signals bond ordering. The Fourier transform
 of bond-bond correlation, the static bond structure factor,  
\begin{equation}
 S_D(q) \equiv \frac{1}{L} \sum_r \cos(qr) D(r),
\label{ftdr}
\end{equation}
identifies the periodicity of the bond modulation. 

We have used DMRG to simulate the Hamiltonian, Eq.(\ref{eq:sp-ham}), on finite-sized chains with periodic boundary conditions.
For each simulation, we have verified that the number of phonon states kept and the number of sweeps applied are sufficient for proper convergence.

For the pure Heisenberg model ($D=0$) coupled to bond phonons, we use a small phonon frequency($\omega=0.05$) to compare our results with the adiabatic limit (classical phonons). Consequently, the average phonon number per bond is expected to be large, and we set the maximum phonon per bond as $n=32$. With a large maximum phonon occupancy, the physical bond dimension for DMRG grows rapidly (for $n=32$, the physical dimension $d=(n+1)\times 3=99$ for each bond). Hence
for the pure Heisenberg model coupled to bond phonons, the largest system size studied is $L=16$,  with maximum phonon per site $n=32$ and maximum states kept in the DMRG calculation $D_{max}=260$;  the maximum number of sweeps is truncated at 100.  

For the model with easy-axis single-ion anisotropy, a larger phonon frequency is used($\omega=1.0$) and the average phonon occupancy is much smaller. Here, we find that when $\alpha=0.6$ and $\omega=1.0$, the maximum phonon per bond $n=4$ is sufficient to characterize the ground state phases. Accordingly, we are able to simulate system sizes up to $L=36$ (with maximum phonon per bond $n=4$ and maximum states kept in DMRG calculation $D_{max}=200$). The maximum number of sweeps is maintained at 100. To properly consider finite-size effects, we have simulated systems with several sizes. 

\section{Results}
The Hamiltonian, Eq.(\ref{eq:sp-ham}), without the spin-phonon interaction ($\alpha = 0$), has been extensively studied using analytical and computational approaches, and the ground state phases in different parameter regimes are well-known\cite{fabricio09}. At zero applied field, the ground state belongs to the celebrated Haldane phase 
for $-0.3 \lesssim D/J \lesssim 0.97$. In this phase, the ground state is a symmetry-protected topological (SPT) state characterized by short-range correlations but exhibiting no long-range magnetic order. For $D/J \lesssim -0.3,$ the ground state has long-range longitudinal antiferromagnetic (L-AF) ordering, while for $D/J \gtrsim 0.97$, the ground state belongs to the quantum paramagnetic phase. All three phases have finite energy gaps to the lowest excitations and the quantum phase transitions are marked by the vanishing of the spin gap. 
\vspace{1.5cm}
\begin{figure}[!ht]
\includegraphics[width=1.0
\columnwidth]{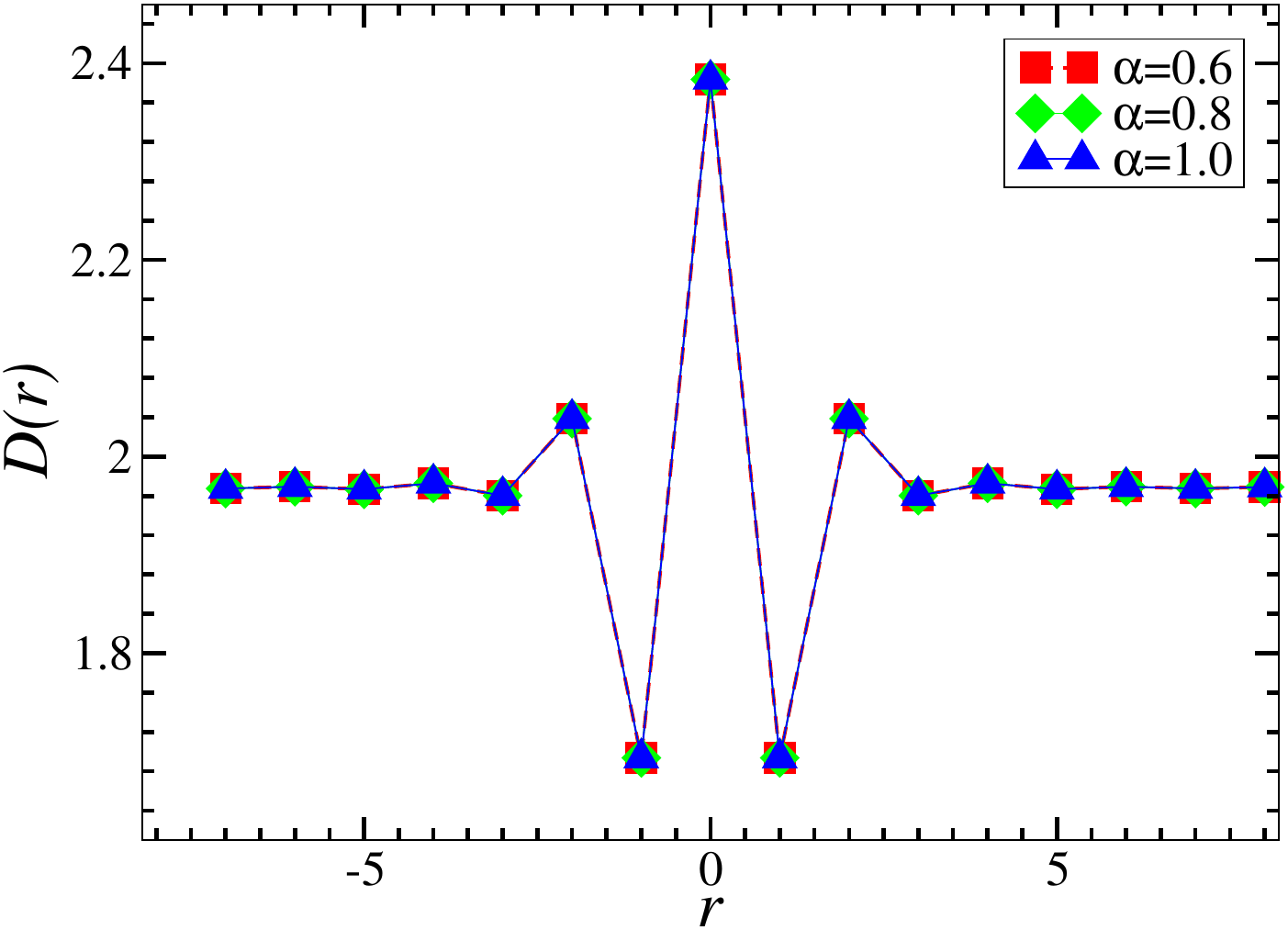}
\caption{(Color online) Bond-bond correlation $D(r)$ as a function of the distance $r$ between two bonds for three representative values of spin-phonon coupling $\alpha$, with $D=h=0$ and $\omega=0.05$, in a system of size $L=16$.} 
\label{fig:spsuppression}
\end{figure}

\subsection{Suppression of Peierls transition due to finite-frequency phonons} 
Now we present our first results. It is well established that when the canonical $S=1$ AFM Heisenberg chain ($D=0$) is coupled to static (classical) bond phonons, there is a spin-Peierls transition to a dimerized state when the spin-phonon coupling strength exceeds $0.25 J$. Our first important finding is that for the pure Heisenberg chain ($D=0$), the spin-Peierls transition is completely suppressed when coupled to finite frequency phonons. The results from the simulation of the Hamiltonian (\ref{eq:sp-ham}) on finite-length chains with $D=h=0$, and fixed phonon frequency $\omega = 0.05$,  are presented in Fig.\ref{fig:spsuppression}. The value of the phonon frequency is deliberately chosen to be small since $\omega\to 0$ represents the classical (static) phonon limit and any spin-Peierls transition is more likely at small $\omega$. 
The figure shows the bond-bond correlation $D(r)$ as a function of the distance $r$ between the bonds for three representative values of (strong) spin-phonon coupling strength, $0.5 \le \alpha \le 1.0$, for system size $L=16$. Any Peierls distortion of the lattice would result in a periodic oscillation of the bond-bond correlation at long distances with a periodicity reflecting that of the lattice distortion. In our simulation results, the bond-bond correlations are oscillatory at short distances (reflecting short-range correlation), but rapidly decay to vanishing values as the inter-bonds distance increases, indicating the absence of any lattice distortion, even for the largest value of $\alpha$. Hence we conclude that quantum lattice fluctuations suppress any 
spin-Peierls transition in the canonical spin-1 AFM Heisenberg chain. While our results cannot rule out a possible transition at even larger values of $\alpha$, such high values are unphysical and not likely to be realized in any real quantum magnet. 

\vspace{0.5cm}
\begin{figure}[!t]
\includegraphics[width=1\columnwidth]{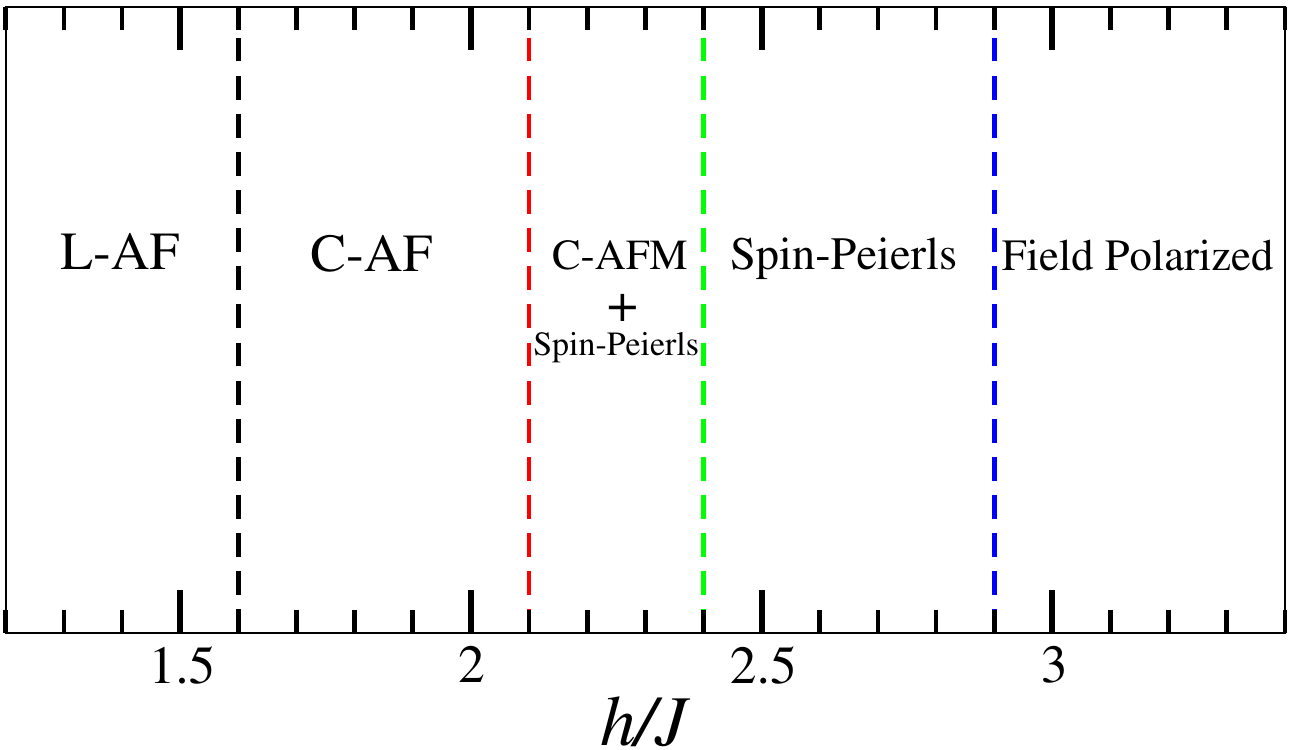}
\caption{(Color online) The phase diagram for $S=1$ Heisenberg chain with strong easy-axis anisotropy $D=-1$ and  spin-phonons coupling $\alpha=0.5$ at finite frequency $\omega=1$. The FISP phase is manifested in the range $2.1 \le h/J \le 2.9$. }  
\label{fig:phase_diagram}
\end{figure}

\subsection{ Field-induced Peierls phase} 
Having established that finite frequency phonons quench any spin-Peierls transition in the canonical Heisenberg model at zero field, we expand our search to other regions of the extended parameter space. 
Our extensive numerical exploration has uncovered a unique spin-Peierls phase that we discuss next. 
In the absence of any spin-phonon coupling ($\alpha=0$), the zero-field ground state of Hamiltonian (\ref{eq:sp-ham}) for easy-axis anisotropy ($D/J \lesssim -0.3$) is in a L-AF phase 
with long-range AFM order and a finite gap to lowest spin excitations. When a longitudinal magnetic field is applied, the ground state spin gap decreases monotonically till it closes and there is a transition of the ground state to a canted AFM (C-AF) phase. In the C-AF phase, the spins are
canted such that the transverse components retain in-plane AFM ordering, whereas the longitudinal components are aligned along the direction of the field. The C-AF phase is marked by a vanishing
spin gap which is reflected in a finite spin stiffness and algebraic decay of spin-spin correlations with distance. 
With increasing field strength, the canting increases continuously in the longitudinal direction until the ground state reaches a fully polarized (FP) phase with all the spins aligned parallel to the applied field. 

\vspace{0.5cm}
\begin{figure}[!htb]
\includegraphics[width=1\columnwidth]{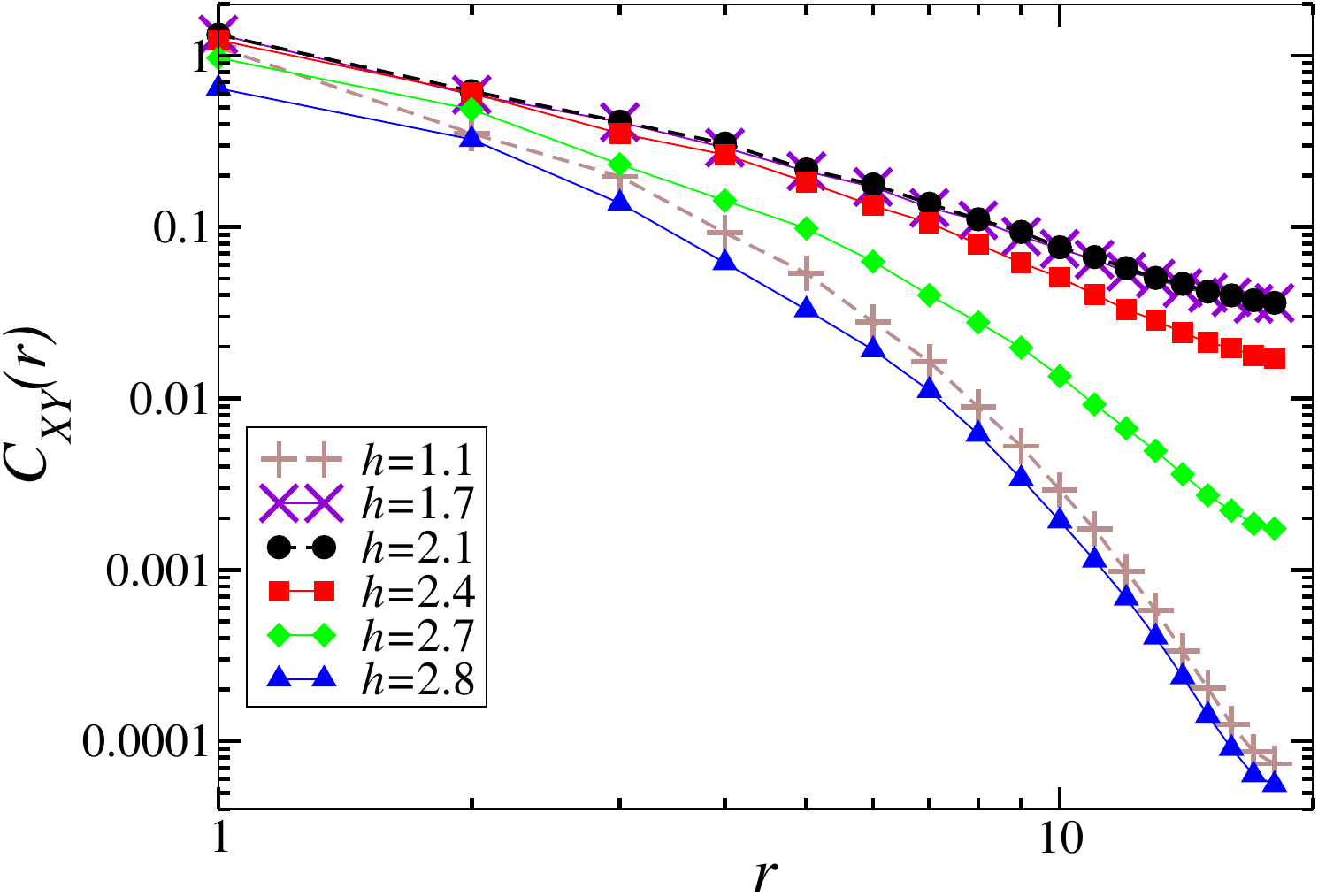}
\caption{(Color online) Green's function $C_{X,Y}(r)$ at different values of $h$ for a system of size $L=36$, with $D=-1$, $\omega=1.0$, and $\alpha=0.5$. }
\label{fig:spin_corr}
\end{figure}

When the coupling between spin degrees of freedom and finite frequency phonons is turned on ($\alpha, \omega \neq 0$), a new field-induced spin-Peierls (FISP) phase emerges. The results of our simulations for a representative set of parameters are summarized in Fig.~\ref{fig:phase_diagram}. For the chosen set of parameters 
($D=-1.0, \omega = 1.0, \alpha=0.5$), the ground state is in the L-AF phase at zero to low field strengths. With increasing $h$, there is a transition out of the 
L-AF phase to the C-AF phase at $h/J\simeq 1.6$, accompanied by the closing of the spin gap. With 
increasing field, there is a transition to a new phase at $h/J \simeq 2.1$ that exhibits the Peierls order (bond strength modulation) while retaining C-AF magnetic ordering. 
Upon further increasing the field strength, the magnetic ordering is quenched at $h/J \approx 2.4$ 
and the ground state exhibits only Peierls order. This is accompanied by the reopening of the spin gap. Finally, at sufficiently strong fields ($h/J \gtrsim 2.9$), the ground state gets field polarized, and the Peierls order is suppressed.    

By carefully characterizing the properties of this new state, we next demonstrate conclusively
that this FISP phase is distinct from any other SP phase previously known. We start with the evolution 
of the spin correlation across the different field-induced phases. Figure \ref{fig:spin_corr} shows 
the transverse component of the equal-time spin-spin correlation as a function of distance for a 
representative set of applied fields in the range $1.1 \le h/J \le 2.8$. Inside the L-AF phase, the 
spin correlation decays exponentially reflecting the absence of any transverse ordering of the 
spins and, equivalently, the presence of a spin gap in the ground state. At intermediate field 
strengths, $1.7 \lesssim h/J \lesssim 2.4$, the transverse spin correlation decays algebraically 
with distance. In this field range, the ground state has C-AF magnetic order, first without ($1.7 
\lesssim h/J \lesssim 2.1$) and then with ($2.1 \lesssim h/J\lesssim 2.4$) co-existing spin-Peierls 
order that will be discussed next. In a one-dimensional system, the algebraic decay of transverse 
spin correlation implies the closing of the spin gap and the presence of quasi-long-range 
transverse magnetic order. 
At higher field strengths ($2.4 \lesssim h/J \lesssim 2.8$), the long-distance behavior of the spin 
correlations changes back to exponential decay, signaling the reopening of the spin gap and the 
suppression of transverse magnetic order. As we shall show next, in this field range, the ground 
state exhibits only spin-Peierls order.

\begin{figure}[!t]
\includegraphics[width=1 \columnwidth]{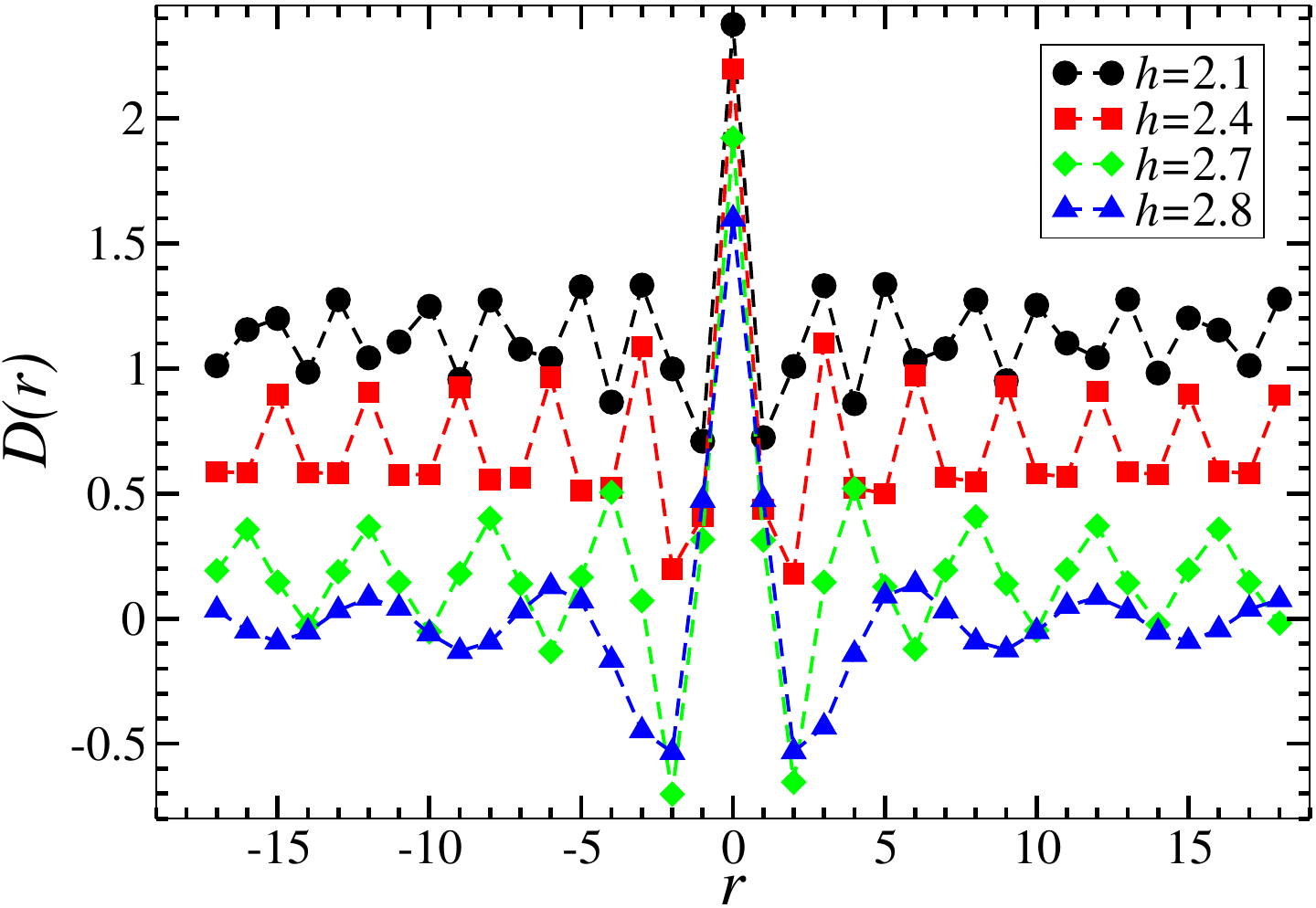}
\caption{(Color online) Bond-bond correlation $D(r)$ at different values of $h$ for a system of size $L=36$, with $D=-1$, $\omega=1$, $\alpha=0.5$.}
\label{fig:dimer_corr}
\end{figure}

To further characterize the putative FISP phase, we turn our attention to the bond-bond correlation $D(r)$ defined in Eq.(\ref{eq:dij}), where a bond is defined as the spin-spin correlation on two neighboring spins. The bond-bond correlation probes the presence of any periodic modulation of bond strengths (equivalently, bond length) across the chain, and, as such, can directly identify an SP state. The results of our simulation are presented in Fig.~\ref{fig:dimer_corr}, where the 
bond-bond correlation is shown for a representative set of magnetic fields in the range $2.1 \lesssim h/J \lesssim 2.9$. These constitute the most important results of this work. The bond-bond correlation is found to exhibit clear signatures of modulation that persist to long distances, indicating the appearance of bond strength modulation, i.e., SP distortion in this phase. 
The strength of modulation is weak close to the transition ($h/J=2.1$) but grows stronger as the system gets deeper into this phase. The bond strength modulation is absent for field strength outside this range, consistent with the predicted nature of the ground state phases shown in
Fig.~\ref{fig:phase_diagram}. 
This is the first instance of the occurrence of spin-Peierls transition driven by quantum phonons in a spin-1 system.  

\begin{figure}[!t]
\includegraphics[width=1
\columnwidth]{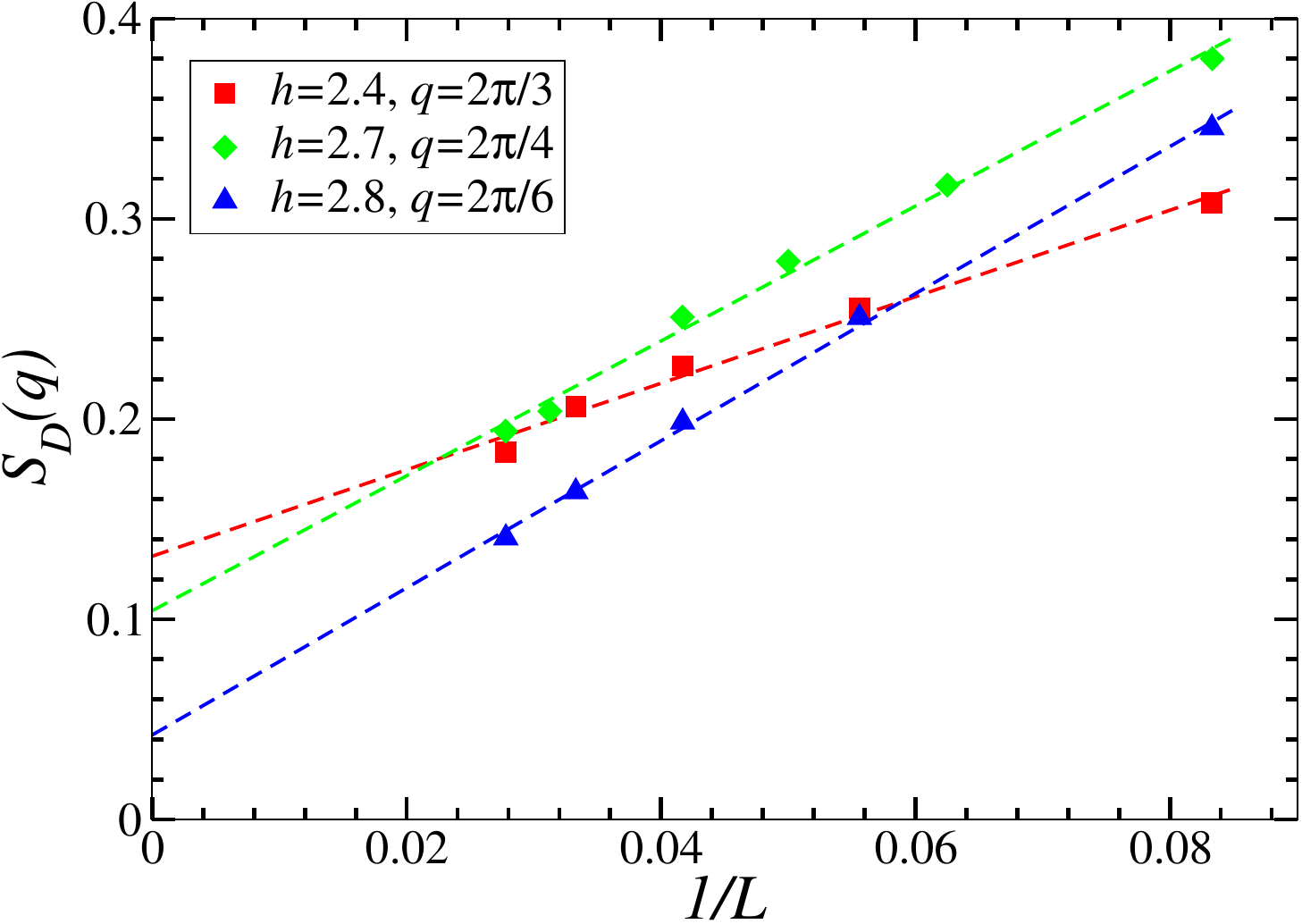}
\caption{(Color online) The static structure factor $S_D(q)$ at the peak momentum $q$ versus $1/L$ for various $h$ values in the FISP phase, with $D=-1$, $\omega=1$, $\alpha=0.5$.}
\label{fig:dk_scaling}
\end{figure}

Most interestingly, the periodicity of modulation in this novel FISP phase is not constant; 
instead, it is governed by the strength of the applied field -- ranging from approximately three lattice sites for $h/J=2.1$ to approximately nine sites for $h/J=2.8$. It is this field-driven emergence and tuning of the periodicity of lattice modulation that sets the present SP phase 
apart from previously reported SP phases. 
To confirm the periodicity of the modulation, we have calculated the static bond structure factor, $S_D(q)$, and carefully analyzed its momentum dependence. Figure \ref{fig:dk_scaling} shows the scaling of the height of the dominant peak with system size for three representative values of the applied field 
inside the SP phase. The peak heights scale to finite, nonzero values, confirming the existence of 
long-range periodicity with the respective values of the crystal momentum.  

\begin{figure}[!htb]
\includegraphics[width=1
\columnwidth]{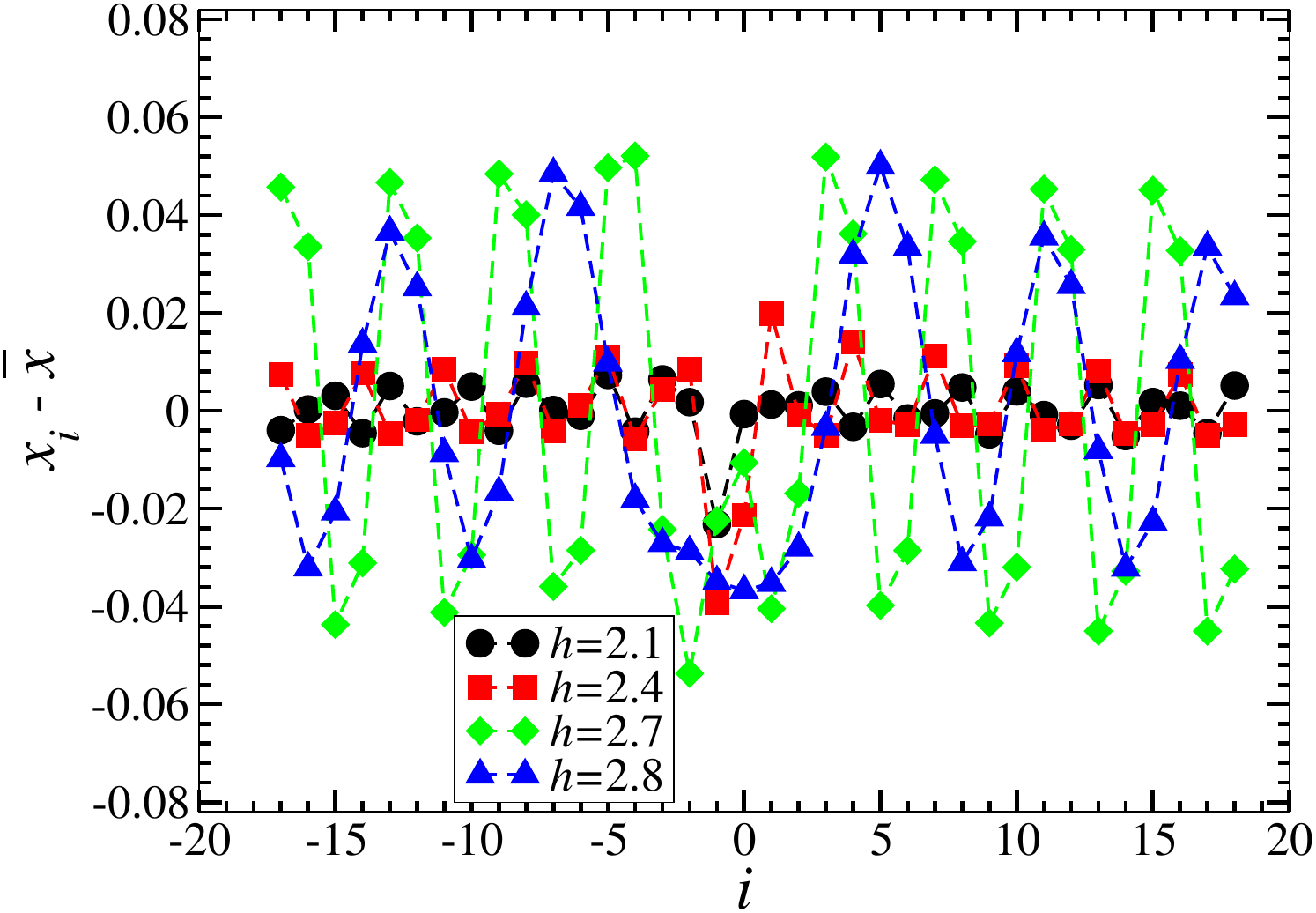}
\caption{(Color online) 
The average phonon displacement $x_i-\Bar{x}$ vs bond index $i$ at different values of $h$ inside the FISP phase for a system size of $L=36$, with $D=-1$, $\omega=1$, $\alpha=0.5$. }
\label{fig:xi}
\end{figure}

To conclude the characterization of the FISP phase, we have calculated the average phonon displacement on individual bonds, defined as 
\begin{equation}
   x_i = \langle {\hat a}_i + {\hat a}_i^\dagger\rangle. 
\end{equation} 
The average phonon displacement over all bonds is $\Bar{x}=\sum_i x_i/L$. For non-zero $x_i$, the bond strength on the bond $i$, between sites $i$ and $i+1$ is modified to $J_i = J+\alpha x_i$. An SP phase is marked by a periodic modulation of $x_i$, and hence $J_i$, which is manifested as a lattice distortion. The results of our simulations are presented in Fig.~\ref{fig:xi}, where the phonon displacement, $x_i-\Bar{x}$, is shown as a function of the bond index for four
representative sets of parameters inside the FISP phase. The data unambiguously show periodic modulation of the average phonon displacement on the bonds. The periodicity of modulation is identical to that of the bond-bond correlation discussed earlier and conclusively identifies the field-induced phase as a spin-Peierls phase.

\vspace{1.0cm}

\section{Conclusion}
To summarize, we have studied the $S=1$ Heisenberg AFM chain, with additional single-ion anisotropy, coupled to finite-frequency bond phonons to search for evidence of spin-Peierls transition in higher spin systems. The results from our large-scale simulations over a wide range
of parameters establish the following: (i) for the canonical Heisenberg chain (vanishing single-ion anisotropy), a transition to the spin-Peierls phase is suppressed for all (realistic values) of the strength of spin-phonon coupling, and, (ii) a unique field-induced Spin-Peierls phase with variable periodicity (that depends on the applied field strength) is realized for easy-axis single-ion anisotropy. 

The complete suppression of SP transition in the presence of finite-frequency phonons contrasts sharply with the results for coupling to static (classical) phonons, for which a transition to the SP phase occurs at a finite critical spin-phonon coupling strength. We conjecture that quantum fluctuations of the phonon field destroy any SP transition for a gapped phase. The ground state of the canonical spin-1 AFM chain is in the Haldane phase, an 
SPT phase with a finite gap. For static phonons, a finite critical spin-phonon coupling—sufficient to close the gap—is required to induce an SP transition, whereas quantum fluctuations of finite-frequency phonons suppress any such transition. 
This is corroborated by our results for easy-axis single ion anisotropy when the ground state is in the gapped L-AF phase in the absence of any applied field and we find no evidence of any SP transition.

%%%%%%%%%%%%%%%%%%%%%%%%%%%%%%%%%%%%%%
\begin{acknowledgments}
S. C. and W. G. were supported by the National Natural Science Foundation of China under Grant No.~12175015. 
P. S. acknowldeges support from the Singapore Ministry of Education through grant MOE-RG159/19(S).
\end{acknowledgments}

%%%%%%%%%%%%%%%%%%%%%%%%%%%%%%%%%%%%%%%%%%%%%%%%%%%%%%%

\end{document}